# Beyond Fermi's Golden Rule in Free-Electron Quantum Electrodynamics: Acceleration/Radiation Correspondence

Yiming Pan[1], Avraham Gover[2]

1. Department of Physics of Complex Systems, Weizmann Institute of Science, Rehovot 76100, ISRAEL
2. Department of Electrical Engineering Physical Electronics, Tel Aviv University, Ramat Aviv 69978, ISRAEL

## Abstract

In this article, we present a unified reciprocal quantum electrodynamics (QED) formulation of quantum light-matter interaction. For electron-light interactions, we bridge the underlying theories of Photon-Induced Near-field Electron Microscopy (PINEM), Laser-induced Particle Accelerators and radiation sources, such as Free Electron Laser (FEL), transition radiation and Smith-Purcell effect. We demonstrate an electron-photon spectral reciprocity relation between the electron energy loss/gain and the radiation spectra. This "Acceleration/Radiation Correspondence" (ARC) conserves the electron-energy and photon-number exchanged and in the case of a Quantum Electron Wavepacket (QEW), displays explicit dependence on the history-dependent phase and shape of the QEW. It originates from an interaction-induced quantum interference term that is usually ignored in Fermi Golden Rule analyses. We apply the general QED formulation to both stimulated interaction and spontaneous emission of classical and quantum light by the quantum-featured electrons. The 'spontaneous' emissions of coherent states ('classical' light) and squeezed states of light are shown to be enhanced with squeezed vacuum. This reciprocal ARC formulation has promise for extension to other fundamental research problems in quantum-light and quantum-matter interactions.

The particle-wave duality of electron and its interaction with light have been a matter of debate and interest since the early days of quantum theory [1-4]. The consistency bridge between the quantum-mechanical wave-like picture of electron-light interaction (e.g. [5-12]) and the classical point-particle picture can be settled by means of a finite size wavepacket presentation of the free-electron [13,14]. In classical theory, the momentum transfer to a point-particle electron, propagating in direction – z, due to interaction with a longitudinal classical electric field component $E = E_{z,cl} \cos(\omega t - q_z z + \phi_0)$ of frequency $\omega$, wavenumber $q_z$ and phase $\phi_0$, is $\Delta p_{point} = -e \int_0^t E_{z,cl} \cos(\omega t - q_z z(t) + \phi_0) dt$. In the linear acceleration approximation (short interaction time, $z(t) - z_0 \simeq v_0 t$), the energy transfer to the accelerated electron is:

$$\Delta E_{point} = v_0 \Delta p_{point} = -e E_{z,cl} L \, \text{sinc}\left(\frac{\bar{\theta}}{2}\right) \cos\left(\frac{\bar{\theta}}{2} + \phi_0\right), \tag{1}$$

with $\bar{\theta} = (\omega/v_0 - q_z)L$ defined as the synchronism detuning parameter and L the interaction length. However, the linear interaction of the same radiation wave in the quantum-mechanical description of the electron as a plane-wave results in a quite different result [5-12].

Recent development in electron microscopy has brought a renewed interest in the quantum analysis of electron-light interaction. Of particular interest is the development of Ultrafast Transmission Electron Microscopy (UTEM) by Zewail and coworkers [7-9] and consequent recent experiments of multi-photon emission/absorption quantum sideband spectra of photon-induced near-field electron microscopy (PINEM) [7,10,11] and wavefunction density attosecond-bunching [15-20]. The underlying theory of PINEM-kind interactions is based on a model of plane-wave-like quantum wavefunction modelling of single electrons. This model results in, for the same configuration of single electron interaction with light, discrete electron energy spectrum of sidebands, spaced apart by the photon energy $\hbar\omega$ .[8,9,21] This spectrum is radically different from the phase-dependent net acceleration energy spectrum of (1), and it cannot be explained in terms of classical point-particle theory. The results of this model are therefore very different from the common formulations of classical free electron interaction schemes with light, such as Free Electron Laser (FEL) [22-25] and other free electron radiation schemes [26-36] and also the recently developing schemes of Dielectric Laser Accelerators (DLA) [19, 37-39].

The bridge between the single electron point-particle acceleration/deceleration model and the PINEM-kind interaction model of electron stimulated interaction with light, is the Quantum Electron Wavepacket (QEW) representation [13,14,40-42]. Based on the solution of Schrodinger equation with a classical radiation field, the linear acceleration (1) of a point-particle is modified in the case of an electron wavepacket to $\Delta E = e^{-\Gamma^2/2} \Delta E_{point}$. Namely, the QEW acceleration is not null, it is phase-dependent, but it is reduced relative to a point particle by a universal factor $e^{-\Gamma^2/2}$, with $\Gamma = 2\pi\sigma_z(t)/\beta\lambda$, where $\beta = v_0/c$ and $\lambda = 2\pi c/\omega$ is the optical wavelength. Therefore, the classical point-particle limit of the accelerated electron wavepacket is given by [13]:

$$\Gamma \ll 1, \text{ and } \sigma_z(t) = \left[\sigma_{z_0}^2 + \frac{1}{4\pi\beta}\left(\frac{\lambda_c^* ct_D}{\sigma_{z_0}}\right)^2\right]^{1/2} \ll \beta\lambda/2\pi. \quad (2)$$

with $\lambda_c^* = \lambda_c/\gamma^3$, $\gamma = (1-\beta^2)^{-1/2}$ the Lorentz factor and $\lambda_c = h/mc$ – the Compton wavelength, and $\sigma_{z_0}$ the intrinsic wavepacket size. The wavepacket size $\sigma_z(t)$ depends on the pre-interaction transport history ($t_D$). It can never be null (ideal 'point-like' particle), because of the Heisenberg uncertainty principle, but in the limit (2), the QEW interacts similarly to a point-like particle (1) compared to the wavelength $\lambda$. In the opposite limit (i.e., $\Gamma \gg 1$), the QEW acceleration is diminished, and the wavepacket-dependent theory predicts the PINEM-kind quantum sidebands energy spectrum. This bridges satisfactorily the two regimes of point particle and quantum wave particle interaction pictures with light in the case of stimulated interaction.

On the other hand, earlier QED analyses of electron-light interaction were carried out only in the intrinsic quantum electron wavefunction limit of an infinitely extended plane-wave electron wavefunction [5,6,12,43,44], and therefore they could not produce the point-particle or QEW acceleration limits. From the first order perturbation QED theory, the spontaneous and stimulated photon emission/absorption rates are derived using Fermi Golden Rule (FGR) [5,6, 40-44]. The net photon emission rate of radiation mode $(q,\sigma)$ at frequency $\omega$, is the difference of the single photon emission and absorption rates: $dv_{q\sigma}/dt = W_{v_f=v_i+1} - W_{v_f=v_i-1}$. For a large class of free electron interaction schemes (e.g., FEL, Smith-Purcell, Cerenkov and Transition Radiations) [5-6,33,43-48], the FGR-based QED model general expression for spontaneous/stimulated photon emission relation is given by

$$\Delta v_q = \Gamma_{sp}\left(\left(v_q(0)+1\right)F\left(\bar{\theta}^{(e)}\right) - v_q(0)F\left(\bar{\theta}^{(a)}\right)\right), \quad (3)$$

where the spontaneous coefficient $\Gamma_{sp}$ and emission/absorption spectral line functions $F\left(\bar{\theta}^{(e,a)}\right)$ depend on the structural light-electron interaction configuration. The stimulated emission is described by the terms with initial photon number $v_q(0) \neq 0$, and the spontaneous emission is described by the only term left when $v_q(0) = 0$. This expression is manifestly different from the semiclassical expression for radiative energy gain and of semiclassical spontaneous emission [20], and unfortunately cannot reduce to the point particle limit of classical radiations [13, 49].

A self-consistent QED model based on a quantum electron wavepacket model and going beyond the FGR approximation [14,20] was shown to result in expressions for spontaneous and stimulated radiation emission that reduce to the point particle limit of electron wavepacket. Here we review briefly this result, and derive independently the expressions for radiation emission and wavepacket acceleration. We show that the QED expressions for the post-interaction electron energy spectrum of a QED reduces to the point particle acceleration/deceleration (1) in the limit (2), and to the PINEM-kind sidebands spectrum in the opposite limit. In this comprehensive analysis, we derive and prove the electron-photon spectral reciprocity and the fundamental energy-conserving Acceleration-Radiation Correspondence (ARC) relation

$$\Delta v_q + \Delta E / \hbar \omega = 0. \quad (4)$$

The reciprocal relation is between the exchanged photon number ($\Delta v_q$) of the radiation field of a single monochromatic mode $(\omega, q)$ and the particle energy transfer (gain/loss, $\Delta E$) of a single electron wavefunction. We present here explicit expressions for spontaneous and stimulated radiative emission and post-interaction energy spectrum of a QEW interacting with a coherent (Glauber) state of the radiation mode that corresponds to the classical electrodynamics limit. We further extend our QED analysis of wavepacket beyond the FGR to cases of electron wavefunction interaction with quantum light sources such as Fock states, coherent states and squeezed states. [50-53]

**The Acceleration/Radiation Correspondence beyond Fermi's Golden Rule** -Let us consider the light-matter interaction of a single electron wavepacket in a general interaction scheme. Single electron wavefunction emission is attainable and well controlled in electron microscopy (i.e., TEM or UTEM) [10, 15, 16], and its interaction with general light pulse can be presented in terms of well-defined photon states. The initial combined electron-photon state can be presented as a superposition of combined electro-photon states in terms of plane-wave representation (momentum p) of the electron wavefunction and Fock photon states of a radiation mode q (suppressed the polarization index for now) in free space [14],

$$|i\rangle = \sum_{p,\nu} c_{p,\nu}^{(0)}(t) |p,\nu\rangle, \qquad (5)$$

where $c_{p,\nu}^{(0)}(t)$ is the component of the combined electron and photon state basis $|p,\nu\rangle$. The electron-photon wavefunction transforms under interaction, in any of the free electron interaction schemes discussed before, exchanging energy and momentum by processes of photon emission/absorption and corresponding electron recoils. The final electron-photon state after interaction is:

$$|f\rangle = \sum_{p,\nu} c_{p,\nu}^{(f)}(t) |p,\nu\rangle, \qquad (6)$$

The light-electron coupling actually redistribute the components of the electron-photon states from $c_{p,\nu}^{(0)} \rightarrow c_{p,\nu}^{(f)}$ within the same representation of electron-photon basis. For our purpose, we describe the coupling of electromagnetic field with the electron, using Schrodinger equation or a 'modified' Klein-Gordon equation (suppressed the spin index) and the interaction Hamiltonian $H_{int} = -e(p \cdot \hat{A} + \hat{A} \cdot p)/m^* + \hat{A}^2/2m^*$ [20,34,44,54]. Note that the first term ($-\hat{A} \cdot p$) applies to "slow wave" light-matter interaction schemes, like Smith-Purcell setup, Cherenkov and transition radiation. The second term describes radiative schemes like Free-Electron Laser, stimulated Compton scattering and Kapitza-Dirac effect [55,56]. In these latter cases, longitudinal force interaction is exerted through the pondermotive potential term ($\hat{A}^2$) [5, 19, 55,56].

In second quantization of the vector potential in terms of photon creation and annihilation operators - $\hat{A} = \hat{A}(a_q^\dagger, a_q)$, we can split the interaction Hamiltonian into two parts: $H_{int} = H_{int}^{(e)} + H_{int}^{(a)}$, in which $H_{int}^{(e,a)}$ correspond to photon emission (e) and absorption (a) respectively. In the weak-coupling perturbation approximation to the order of single-photon emission/absorption, we describe these processes in terms of reciprocal scattering ladders representing the electron acceleration and photon radiation, as depicted in Fig.1.

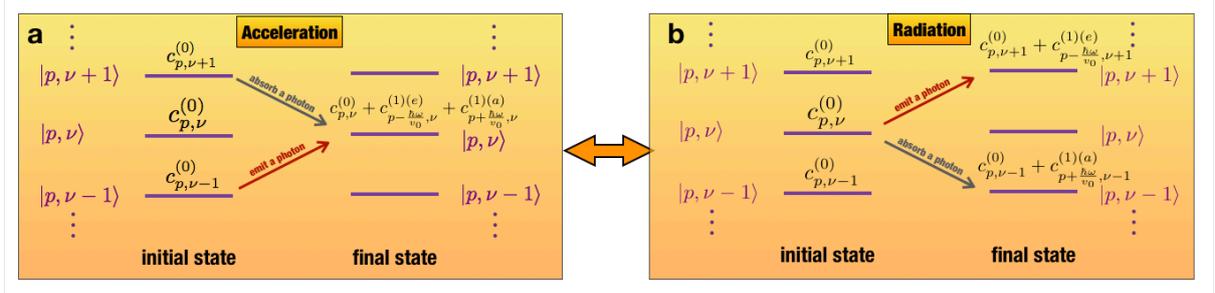

*Fig. 1: The correspondence of electron acceleration (left, a) and photon radiation (right, b) in quantum light-matter interaction schemes.*

In a very general universal description of free-electron interaction schemes with quantum light, these diagrams manifest implicitly the Acceleration-Radiation Correspondence (ARC) (4). Explicitly, the *rhs* diagram of Fig. 1 indicates, that by emitting a photon, the final coefficient of state $|p, \nu+1\rangle$ is given by $c_{p,\nu+1}^{(0)} + c_{p-\frac{\hbar\omega}{v_0},\nu+1}^{(1)(e)}$, where $c_{p,\nu+1}^{(0)}$ is the initial coefficient, and $c_{p-\frac{\hbar\omega}{v_0},\nu+1}^{(1)(e)}$ represents a reciprocal electron momentum (or energy) conserving process of photon emission and momentum backward recoil: $|p,\nu\rangle \Rightarrow |p-\frac{\hbar\omega}{v_0}, \nu+1\rangle$. Likewise, by absorbing a photon, the final coefficient of state $|p,\nu-1\rangle$ is given by $c_{p,\nu-1}^{(0)} + c_{p+\frac{\hbar\omega}{v_0},\nu-1}^{(1)(a)}$, corresponds to absorption of a photon and electron momentum forward recoil: $|p,\nu\rangle \Rightarrow |p+\frac{\hbar\omega}{v_0}, \nu-1\rangle$. Therefore, describing the physical consequence of the stimulated photon emission as the imbalance between contributions of emission and absorption processes, results in (Fig.1a):

$$\Delta\nu_q = \sum_{p,\nu} \left| c_{p,\nu+1}^{(0)} + c_{p-\frac{\hbar\omega}{v_0},\nu+1}^{(1)(e)} \right|^2 - \left| c_{p,\nu-1}^{(0)}(t) + c_{p+\frac{\hbar\omega}{v_0},\nu-1}^{(1)(a)} \right|^2, \qquad (7)$$

The scattered components $c^{(1)(e,a)}_{p\mp(\hbar\omega/v_0),v\pm1}$ represent the reciprocal momentum conserving processes through emission or absorption of a single photon and momentum recoils, relating to the initial component $c^{(0)}_{p,v}$ and also a specific photon-scattering matrix elements $\left|\langle f|H^{(e,a)}_{int}|i\rangle\right|$.

On the other hand, minding the electron spectral energy/momentum transfer on the *lhs* of Fig. 1, the particle acceleration of electron wavefunction in the single-photon emission and absorption processes is given by

$$\Delta E = \sum_{p,v}\left|c^{(0)}_{p,v} + c^{(1)(e)}_{p-\frac{\hbar\omega}{v_0},v} + c^{(1)(a)}_{p+\frac{\hbar\omega}{v_0},v}\right|^2 (E_p - E_0), \qquad (8)$$

where $E_p$ is the electron energy-momentum dispersion relation: $E_p = E_0 + v_0(p-p_0) + (p-p_0)^2/2m^*$, where $m^* = \gamma_0^3 m$, and $E_0 = \sum_{p,v}\left|c^{(0)}_{p,v}\right|^2 E_p$ is the initial electron energy and $p_0$ is the average momentum. Note that the final state $c^{(f)}_{p,v} = c^{(0)}_{p,v} + c^{(1)(e)}_{p-\frac{\hbar\omega}{v_0},v} + c^{(1)(a)}_{p+\frac{\hbar\omega}{v_0},v}$ includes the photon-emitted (e) and photon-absorbed (a) contributions in terms of the final scattered redistribution: $|p,v\mp1\rangle \Rightarrow |p\mp\hbar\omega/v_0,v\rangle$.

To prove the general correspondence of radiation and acceleration (ARC) (4), we expand the expressions for the photon emission and electron acceleration $\Delta v_q, \Delta E$, respectively, cancel the initial terms, and rewrite separately the phase-dependent interference terms of the squared binomials $\Delta v_q^{(1)}, \Delta E^{(1)}$ and the phase-independent scattering terms $\Delta v_q^{(2)}, \Delta E^{(2)}$:

$$\begin{aligned}\Delta v_q^{(1)} &= 2\sum_{p,v}\Re\left\{\left(c^{(0)*}_{p,v+1}\cdot c^{(1)(e)}_{p-\frac{\hbar\omega}{v_0},v+1}\right) - \left(c^{(0)*}_{p,v-1}\cdot c^{(1)(a)}_{p+\frac{\hbar\omega}{v_0},v-1}\right)\right\}, \\ \Delta E^{(1)} &= 2\sum_{p,v}\Re\left\{c^{(0)*}_{p,v}\cdot c^{(1)(e)}_{p-\frac{\hbar\omega}{v_0},v} + c^{(0)*}_{p,v}\cdot c^{(1)(a)}_{p+\frac{\hbar\omega}{v_0},v}\right\}(E_p - E_0),\end{aligned} \qquad (9)$$

and

$$\begin{aligned}\Delta v_q^{(2)} &= \sum_{p,v}\left|c^{(1)(e)}_{p-\frac{\hbar\omega}{v_0},v+1}\right|^2 - \left|c^{(1)(a)}_{p+\frac{\hbar\omega}{v_0},v-1}\right|^2, \\ \Delta E^{(2)} &= \sum_{p,v}\left|c^{(1)(e)}_{p-\frac{\hbar\omega}{v_0},v} + c^{(1)(a)}_{p+\frac{\hbar\omega}{v_0},v}\right|^2 (E_p - E_0),\end{aligned} \qquad (10)$$

Keeping the dispersion relation approximation with respect to momentum p to the first order, one substitutes separately in the sums (9, 10) $E_p - E_0 = \hbar\omega$ for the absorption terms and $E_p - E_0 = -\hbar\omega$ for the emission terms. This reverses the signs between the terms and establishes the ARC relation for each order

$$\Delta v_q^{(1,2)} + \Delta E^{(1,2)}/\hbar\omega = 0, \quad (11)$$

as well as for the general relation (4). Note that the phase-dependent emission/absorption and acceleration/deceleration terms (9) are dependent on the radiation light phase through the scattering matrix element, but the phase-independent terms (10) are not. In conventional QED analysis, the photon emission rate is calculated in perturbation theory based on Fermi's golden rule (FGR) [5,6,34,40-46], keeping the phase-independent term only (10) and neglecting the mixture terms, thus, missing the interesting physics of interference between overlapping initial and scattered terms. Keeping the contribution of interference, beyond the FGR model, is a pivotal methodological step in our formulation in the present article, and also an inevitable necessity for bridging the quantum plane-wave and classical point-particle description of the electron wavefunction, which is the essence of wave-particle duality [13,14,20].

We also note that in the present analysis we neglect the contribution of possible interference between the emission and absorption scattering amplitudes

$$\sum_{p,v} 2\Re\left\{c_{p-\frac{\hbar\omega}{v_0},v}^{(1)(e)} \cdot c_{p+\frac{\hbar\omega}{v_0},v}^{(1)(a)}\right\}. \quad (12)$$

This emission-absorption interference term is small in the first order perturbation analysis when $\left|c_{p,v}^{(0)}\right|^2 \gg \left|c_{p\mp\frac{\hbar\omega}{v_0},v\pm1}^{(1)(e,a)}\right|^2$. We thus assume $\Delta E^{(1)} \gg 2\sum_{p,v}\Re\left\{c_{p-\frac{\hbar\omega}{v_0},v}^{(1)(e)} \cdot c_{p+\frac{\hbar\omega}{v_0},v}^{(1)(a)}\right\}(E_p - E_0)$, and concentrate on the interference term between initial and scattered components, neglecting, to first order, the higher-order interference between emission and absorption components in the first-order perturbation approximation. However, we point out that this interference term may be important in shaping the electron wavefunction in the case $\left|c_{p,v}^{(0)}\right|^2 \ll \left|c_{p\mp\frac{\hbar\omega}{v_0},v\pm1}^{(1)(e,a)}\right|^2$, which may relate to the case discussed by Murdia and coauthors [57] in the context of Bremsstrahlung.

**Electron-photon interaction in a slow-wave structure (tip/grating/foil)** - The formulation, so far, is valid for the wide class of electron-photon interaction schemes reviewed in [5,34,47]. They only differ in the derivation of the interaction matrix elements. Following [13,14], we exemplify here the explicit derivation for the interaction schemes, in which a slow-wave axial field component of a radiation mode $(\omega, q_z)$ interacts synchronously with a co-propagating electron wavefunction of velocity $v_0$ along an extended interaction length. Neglecting the ponderomotive potential term ($\hat{\mathbf{A}}^2$), the interaction Hamiltonian in minimal coupling is taken to be

$$H_I(t) = -\frac{e\left(\hat{\mathbf{A}}\cdot(-i\hbar\nabla)+(-i\hbar\nabla)\cdot\hat{\mathbf{A}}\right)}{2\gamma_0 m}. \tag{13}$$

For the case of our concern $\hat{\mathbf{A}} = -\frac{1}{2i\omega}\left(\hat{\mathbf{E}}(\mathbf{r})e^{-i\omega t} - \hat{\mathbf{E}}^\dagger(\mathbf{r})e^{i\omega t}\right)$, where $\hat{\mathbf{A}}, \hat{\mathbf{E}}$ are vector potential and electrical field operators. In our one-dimensional analysis we assume for simplicity, that the light-electron coupling takes place through an axial slow-wave field component of one of the modes q of a quantized field radiation mode expansion [34]: $\hat{\mathbf{E}}(\mathbf{r}) = \sum_q \tilde{E}_{qz} e^{iq_z z - i\phi_0} \hat{a}_q \mathbf{e}_z$, where $\hat{a}_q (\hat{a}_q^\dagger)$ is the annihilation (creation) operator of photon number state $|v\rangle$. For the example of Cerenkov radiation [30, 33], $E_{qz} = E_{\perp z}\sin\Theta$ and $q_z = n(\omega)(\omega/c)\cos\Theta$, where $\Theta$ is the wave propagation angle relative to the electron propagation axis z, $E_{q\perp}$ is the transverse component of the wave, and $n(\omega)$ is the index of refraction of the medium. For Smith-Purcell setup, the near-field is the axial component of one of the space-harmonics of a Floquet-mode radiation wave incident on a grating [13, 20]: $\hat{E}_q(\mathbf{r}) = \sum_m \tilde{E}_{qzm} e^{iq_{zm}z}$, where $q_{zm} = q_{z0} + m2\pi/\lambda_G$, and one of the space-harmonics, with $q_z = q_{zm}$ satisfies near synchronism condition selectively: $\omega/q_z \simeq v_0$ [14].

The first-order time-dependent perturbation theory, then results in $i\hbar \dot{c}_{p',v'}^{(1)} = \int \frac{dp}{\sqrt{2\pi\hbar}} \sum_v c_{p,v}^{(0)} \langle p',v'|H_I(t)|p,v\rangle e^{-i(E_p - E_{p'})t/\hbar}$. Integrating in time from 0 to infinity, the emission and absorption terms of the perturbation coefficients $c_{p',v'}^{(1)} = c_{p',v'}^{(1)(e)} + c_{p',v'}^{(1)(a)}$ are given by

$$c_{p',v'}^{(1)(e,a)} = \begin{cases} +\left(\dfrac{p'+p_{rec}^{(e)}-\hbar q_z/2}{p_0}\right)\tilde{\Upsilon}\sqrt{v'}\,c_{p'+p_{rec}^{(e)},v'-1}^{(0)}\operatorname{sinc}\left(\overline{\theta}^{(e)}/2\right)e^{i\left(\overline{\theta}^{(e)}/2+\phi_0\right)}, \\ -\left(\dfrac{p'-p_{rec}^{(a)}+\hbar q_z/2}{p_0}\right)\tilde{\Upsilon}\sqrt{v'+1}\,c_{p'-p_{rec}^{(a)},v'+1}^{(0)}\operatorname{sinc}\left(\overline{\theta}^{(a)}/2\right)e^{-i\left(\overline{\theta}^{(a)}/2+\phi_0\right)}, \end{cases} \quad (14)$$

with the normalized photon exchange coefficient $\tilde{\Upsilon}=e\tilde{E}_{qz}L/4\hbar\omega$, and $\overline{\theta}^{(e,a)}=\left(p_{rec}^{(e,a)}\pm\hbar q_z\right)L/\hbar=\overline{\theta}\pm\varepsilon/2$, where $\overline{\theta}=\left(\omega/v_0-q_z\right)L$ is the classical interaction 'detuning parameter', $\varepsilon=\delta\left(\omega/v_0\right)L$ is the negligible interaction quantum recoil parameter. [5,13].

The photon emission and electron energy transfer of ARC (4) can be calculated for a Fock state initial condition by just substituting $c_{p,v}^{(0)}=c_p^{(0)}\delta_{v,v_0}$ in (14), and using it in Eqs (9-10) for the photon and electron energy increments. With the reasonable approximations $p_{rec}^{(e,a)}/p_0\ll 1$, $\hbar q_{zm}/p_0\ll 1$, this results in [14] for the phase-independent term

$$\Delta v_q^{(2)}=-\Delta E^{(2)}/\hbar\omega=\tilde{\Upsilon}^2\left\{(v_0+1)\operatorname{sinc}^2\left(\overline{\theta}^{(e)}/2\right)-v_0\operatorname{sinc}^2\left(\overline{\theta}^{(a)}/2\right)\right\}. \quad (15)$$

However, the phase-dependent terms in (9,10) are null: $\Delta v_q^{(1)}=-\Delta E^{(1)}/\hbar\omega=0$. Namely, there is no phase-dependent contribution of a single Fock state $|v_0\rangle$, neither for $v_0\neq 0$, nor for $v_0=0$. The conclusion is that there is no phase-dependent spontaneous emission in vacuum from a single electron wavepacket, and the only QED contribution to spontaneous emission is phase-independent (resulting from the "zero-point vibration" [5,14]):

$$\Delta v_{q,\text{vac}}=\tilde{\Upsilon}^2\operatorname{sinc}^2\left(\overline{\theta}/2\right), \quad (16)$$

These results are identical with the stimulated and spontaneous emissions derived before for a plane wave quantum wavefunction model of the electron (3) [5,6], and unfortunately, they cannot provide the transition to the classical point particle picture (1) of electron-light interaction.

**Interaction of Quantum Electron Wavepacket with Coherent State Light and its Quantum to Classical Transition** - Consider the case of an initial state of electron wavepacket, modeled as a chirped Gaussian distribution combined and a coherent photon state ($\left|\sqrt{v_0}\right\rangle$)

$$c_{p,v}^{(0)} = c_p^{(0)} c_v^{(0)}, \tag{17}$$

with

$$c_p^{(0)} = \left(2\pi\sigma_{p_0}^2\right)^{-1/4} \exp\left(-\frac{(p-p_0)^2}{4\tilde{\sigma}_p^2(t_D)}\right) e^{i(p_0 L_D - E_0 t_D)/\hbar},$$

$$c_v^{(0)} = e^{-v_0/2} \sum_{v=0}^{\infty} \frac{(v_0)^{v/2}}{\sqrt{v!}}, \tag{18}$$

where $\tilde{\sigma}_p^2(t_D) = \sigma_{p_0}^2 \left(1 + i\xi t_D\right)^{-1}, \xi = 2\sigma_{p_0}^2/m^*\hbar, L_D = v_0 t_D$ and $v_0$ is the expectation value of the photon number of mode q. Substituting (17-18, 14) into the formulae (9-10), one can check the electron acceleration and photon radiation independently, and thus obtain explicitly the ARC relation

$$\Delta v_q^{(1)} = -\Delta E^{(1)}/\hbar\omega = \left(\frac{eE_{z,cl}L}{\hbar\omega}\right) e^{-\Gamma^2/2} \text{sinc}\left(\frac{\bar{\theta}}{2}\right)\cos\left(\frac{\bar{\theta}}{2} + \phi_0\right),$$

$$\Delta v_q^{(2)} = -\Delta E^{(2)}/\hbar\omega = \tilde{\Upsilon}^2 \left\{(v_0 + 1)\text{sinc}^2\left(\bar{\theta}^{(e)}/2\right) - v_0 \text{sinc}^2\left(\bar{\theta}^{(a)}/2\right)\right\}. \tag{19}$$

Note that $v_0$ in this equation is the photon number expectation value of the coherent state (18) replacing the Fock state number in (16). Here we used the following relations

$$\left\langle\sqrt{v_0}\left|\hat{a}_q\right|\sqrt{v_0}\right\rangle = \sqrt{v_0}, \quad \left\langle\sqrt{v_0}\left|\hat{a}_q^\dagger\right|\sqrt{v_0}\right\rangle = \sqrt{v_0},$$

$$\left\langle\sqrt{v_0}\left|\hat{a}_q^\dagger\hat{a}_q\right|\sqrt{v_0}\right\rangle = v_0, \quad \left\langle\sqrt{v_0}\left|\hat{a}_q\hat{a}_q^\dagger\right|\sqrt{v_0}\right\rangle = v_0 + 1, \tag{20}$$

for a coherent state and $\sqrt{v_0}\tilde{E}_{qz} = E_{z,cl}$, where $E_{z,cl}$ is the axial slow wave field component of the interacting radiation mode (1).

In the limit of recoil parameter $\varepsilon \ll 1$, the phase-independent ARC expression (19b) reduces to the famous low-gain expression of FEL [5,13,23,24]:

$\Delta v_q^{(2)} = -\Delta E^{(2)}/\hbar\omega = \tilde{\Upsilon}^2 \left\{ \text{sinc}^2(\bar{\theta}/2) + v_0 \varepsilon \, d\left(\text{sinc}^2(\bar{\theta}/2)\right)/d\bar{\theta} \right\}$. The ever-existing spontaneous emission contribution (16) of the zero-point quantum vibration of Fock state $v_0 = 0$ is kept in (19b). The phase-dependent interference term (19a), results in the semiclassical expression for stimulated interaction with a QEW [14,35,36], including the significant extinction coefficient $e^{-\Gamma^2/2}$, where

$$\Gamma = \left(\frac{\omega}{v_0}\right)\sigma_z(t_D) = \left(\frac{\hbar\omega}{v_0}\right)\frac{\sqrt{1+\xi^2 t_D^2}}{2\sigma_{p_0}} = \Gamma_0\sqrt{1+\xi^2 t_D^2}, \quad (21)$$

with $\Gamma_0 = 2\pi\sigma_{z_0}/\beta\lambda$. This "beyond FGR" relation for wavepacket-dependent acceleration (19a) results out of the commonly neglected interference terms (9) after integration with the center-shifted Gaussian distributions (18) (see Appendix A). It confirms the ARC relation (4) and demonstrates the transition from the quantum electron plane-wave limit $\Gamma \gg 1$, where only phase-dependent emission/acceleration exists, to the point-particle limit of classical linear acceleration (1) in the opposite limit (2). Moreover, it also demonstrates the emission/acceleration dependence on the history-dependent wavepacket size (21) in the quantum to classical transition regime.

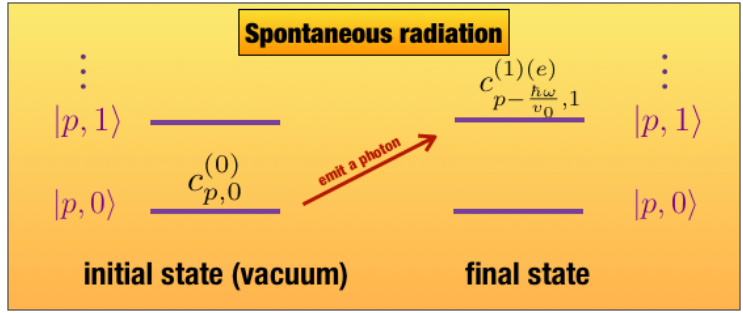

*Fig. 2: Spontaneous emission (quantum noise) occurs due to the incident radiation field at vacuum state. In the absence of any light sources, there is no process of photon absorption.*

Confirmation of the ARC relation and the wavepacket dependence prediction of (19a), The observation of the radiation phase-dependent and wavepacket size-dependent term (19a) would be hampered in practice by the phase-independent term (19b). Even in the case $\bar{\theta} = 0$, when the second order stimulated emission part vanishes, the spontaneous emission part (16) ever persists. Thus, it acts as background quantum noise, such that the detection of the phase dependent term requires exceeding a signal to noise limit condition

$$\frac{S}{N} \equiv \left(\frac{\Delta v_q^{(1)}}{\Delta v_{q,\text{vac}}}\right)\bigg|_{\max} = \frac{4\sqrt{v_0}}{\tilde{\Upsilon}}. \tag{22}$$

Thus, the signal to noise condition (S/N>1) for detection of the phase-dependent signal requires sufficiently intense incident laser pulse $v_0 > \tilde{\Upsilon}^2/16$. The explicit determination of the quantum spontaneous emission per mode depends on the configuration of the radiative interaction scheme. In a model of a radiation plane wave incident on a slow wave structure of length L (a Cerenkov medium or a Smith-Purcell structure), one may quantize the travelling wave radiation mode by the single mode energy quantization requirement $\frac{1}{2}\sqrt{\frac{\varepsilon_0}{\mu_0}}\left|\tilde{E}_{q\perp 0}\right|^2 A_{\text{eff}} t = \hbar\omega$, where $t = L/v_0 = L/\beta c$ is the interaction time along the structure and $A_{\text{eff}}$ is the effective cross-section of a diffraction limited incident radiation mode. This model approximation was shown good enough to result in expressions for spontaneous emission of Smith-Purcell radiation in an open structure, consistent with independent semiclassical and QED derivations [5, 20, 44, 54] (see Appendix B).

Summarizing this section of quantum electron wavepacket interaction with classical light, we determined that the transition from the quantum plane wave limit of the electron to the classical 'point-particle' limit requires that the radiation wave is a classical coherent state of light, and the extinction coefficient in (19a) $\Delta E^{(1)} = \Delta E_c e^{-\Gamma^2/2}, \Delta v_q^{(1)} = \Delta v_c e^{-\Gamma^2/2}$ tends to $e^{-\Gamma^2/2} \to 1$. This requires that the electron wavefunction spatial distribution is point-like, with the wavepacket-size satisfying condition: $\Gamma \ll 1$ ($\sigma_z \ll \beta\lambda/2\pi$). In the opposite limit ($\Gamma \gg 1$) the phase dependent term (19a) diminishes: $\Delta E^{(1)} \to 0, \Delta v_q^{(1)} \to 0$, and only the phase-independent term (19b) remains, which defines the earlier known quantum plane-wave limit [5]. Fig. 3a depicts the classical to quantum wavepacket transition of the new "beyond FGR" first order interference term (19a). Besides the wavepacket-dependent Gaussian decay, it shows in broken line also the non-vanishing wavepacket-independent spontaneous emission contribution of the second order term (16) that acts as noise, and must be exceeded in order to observe the first order stimulated emission. Fig. 3b is a color code presentation of the extinction coefficient $e^{-\Gamma^2/2}$ as a function of the intrinsic wavepacket size (at its waist) $\Gamma_0 = 2\pi\sigma_{z_0}/\beta\lambda$, and the drift length of the wavepacket $L_D=v_0 t_D$ before the interaction point. It demonstrated that the phase-dependent interaction of the quantum wavepacket depends, curiously enough, on its history before interaction.

Experimental confirmation of the ARC relation and the wavepacket dependence prediction of (19a), would require simultaneous measurement of photon emission and electron energy spectra of single electron radiative interaction events. This is a challenging experiment that requires accumulating multiple interaction events data of electron wavepacket radiative interaction and preselection of the wavepackets, using wavepacket shape-formation schemes [15-17] as suggested in ref. [58]. The detection of the radiation certainly requires also satisfaction of the signal-to-noise ratio condition S/N>1, considering the ever present wavepacket-independent noise due to spontaneous emission (quantum noise).

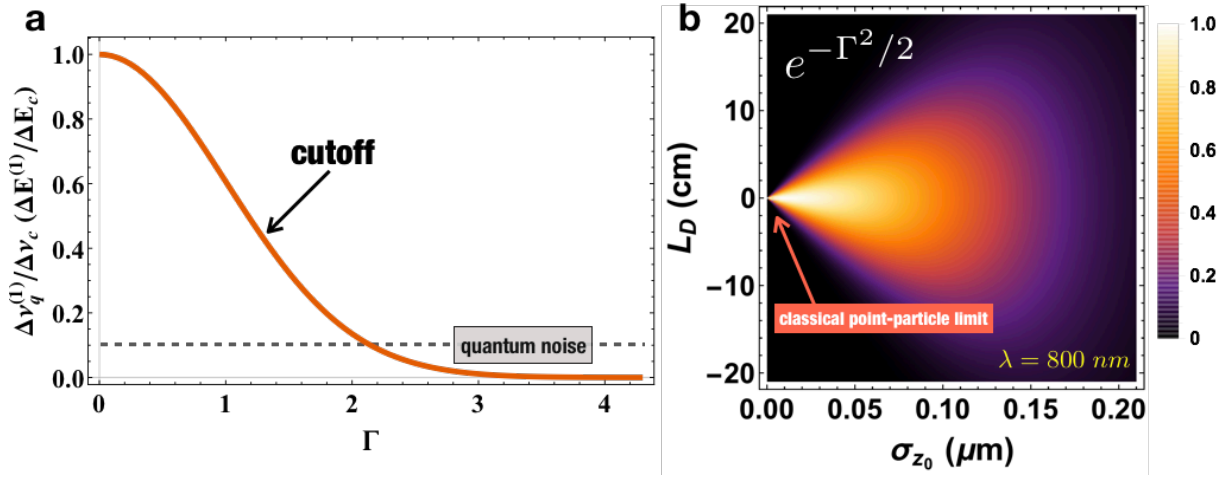

*Fig. 3: (a)The wavepacket-dependent photon emission rate as a function of the decay factor $\Gamma = 2\pi\sigma_z(L_D)/\beta\lambda$ for electron wavepacket interacting with a coherent laser beam. The spontaneous emission quantum noise is schematically depicted in broken line for comparison with the phase-dependent stimulated emission curve for case of signal to noise ratio (S/N>1). (b)The history and wavepacket-size dependence of the acceleration/radiation extinction coefficient $e^{-\Gamma^2/2}$ with $\Gamma = \Gamma(\sigma_{z_0}, L_D)$ defined in Eq. 18 of the text. $L_D <0$ corresponds to the case where the negative chirped wavepacket converges to its waist point beyond the start interaction point (drawn for $\lambda = 800\,\text{nm}, \beta = 0.7$). The classical point-particle interaction of electron is in the limit of $\Gamma \to 0$.*

**The photon 'size' effect of quantum light - squeezed coherent state** - Coherent and squeezed states of photons are two basic elements of quantum optics. Coherent state is the closest analog to a classical electromagnetic field, but squeezed state has no such classical counterpart. [50,51,53]. The phase squeezing of photon coherent state in quadrature representation is analogous to the control over the size of the chirped Gaussian quantum electron wavepacket in phase-space in the present work. It is thus interesting to examine the 'photon size' effect of squeezed light, when interacting with a finite size quantum electron wavepacket. Following the standard definition [50], a squeezed coherent state is generated by applying on the vacuum state, the displacement operator $D(\alpha) = \exp(-\alpha\hat{a}_q^\dagger - \alpha^*\hat{a}_q)$, followed by the squeezing operator $S(\xi) = \exp(\xi^*\hat{a}_q^2/2 - \xi\hat{a}_q^{\dagger 2}/2)$, i.e.,

$$|\xi, \alpha\rangle = S(\xi)D(\alpha)|0\rangle, \tag{23}$$

where $|\alpha|^2 = \nu_0$, and $\xi$ is the squeezing parameter of the unitary squeezing operator $S(\xi)$.

We can now derive the radiation/acceleration correspondence (11) for this case, starting from (9-10), following the derivation of (19) in the coherent state case, but using now in the calculation of the matrix elements the expectation values of squeezed coherent photon state [40-42, 60]:

$$\langle\xi, \alpha|\hat{a}_q|\xi, \alpha\rangle = \sqrt{\nu_0}, \quad \langle\xi, \alpha|\hat{a}_q^\dagger|\xi, \alpha\rangle = \sqrt{\nu_0},$$
$$\langle\xi, \alpha|\hat{a}_q^\dagger\hat{a}_q|\xi, \alpha\rangle = \nu_0 + \sinh^2|\xi|. \tag{24}$$

The first two expressions for the expectation value of field strength are not different from the first two expressions in (20) for the case of coherent state of light, and thus do not change the result for the phase-dependent ARC relation (19a). However, the third expression in (24) makes a difference in the calculation of the matrix element of the phase-independent term (19b). Therefore, the ARC expression for the squeezed state case is modified to

$$\Delta\nu_q^{(1)} = -\Delta E^{(1)}/\hbar\omega = 4\tilde{\Upsilon}\sqrt{\nu_0}e^{-\Gamma^2/2}\text{sinc}\left(\frac{\bar{\theta}}{2}\right)\cos\left(\frac{\bar{\theta}}{2} + \phi_0\right),$$
$$\Delta\nu_q^{(2)} = -\Delta E^{(2)}/\hbar\omega = \tilde{\Upsilon}^2\left\{(\nu_0 + \sinh^2|\xi| + 1)\text{sinc}^2\left(\bar{\theta}^{(e)}/2\right) - (\nu_0 + \sinh^2|\xi|)\text{sinc}^2\left(\bar{\theta}^{(a)}/2\right)\right\}.$$
$$(25)$$

The coherent state is position-squeezed at $\xi > 0$, and momentum squeezed at $\xi < 0$, where the relative photon 'size' is defined as $e^{-|\xi|}$ in the dimensionless quadratures. We have found that the squeezing has no effect on the ARC relation of the wavepacket phase-dependent interaction, and certainly not on its point particle limit. In second order, we have found such dependence on the absolute squeezing factor $|\xi|$, increasing the effective input photon number to ($v_0 + \sinh^2|\xi|$). In particular, the conventional spontaneous emission expression (16) is modified for the case of squeezed vacuum state ($v_0 = 0$), and is given in the limit of negligible interaction recoil $\varepsilon \ll 1$, by

$$\Delta v_{q,squ} = \Delta v_{q,vac} + \tilde{\Upsilon}^2 \sinh^2|\xi| \left( \varepsilon \frac{d}{d\bar{\theta}} \text{sinc}^2 \left( \frac{\bar{\theta}}{2} \right) \right), \qquad (26)$$

which can be interpreted (compare to (19)) as FEL-kind amplification of the squeezed vacuum state. It is worth noting that this interesting new term is similar to the Hawking-Unruh effect for accelerating observers (photon detection), where the Unruh-like temperature (T) for our case is analogously defined by the squeezing photon number $\sinh^2|\xi| = \left( e^{\hbar\omega/k_B T} - 1 \right)^{-1}$. [53,60-62] The primary connection is that the squeezed operator S, mathematically shares the Bogoliubov transformation from the Kindler space-time of non-inertial observer to Minkowski space-time, especially for the case of squeezed states of quantum light. [53, 55-56] As a result, we should note that the squeezed radiation emission expression we found (25b), is a similar radiative process analogic to the Unruh effect.

Summarizing the results of this section on QEW interaction with squeezed quantum light, we have not found any quantum photon 'size' effect in connection to wavepacket size and phase dependent interaction term. However, we find that the phase-independent acceleration/emission term of a QEW with squeezed light state of finite photon size ($e^{-|\xi|}$), exhibits non-classical optical properties of non-Poisson distributions. Thus, the general theoretical framework of ARC spectral reciprocity presented here lays ground for studying super- or sub-Poissonian photon distributions in numerous classical or quantum light sources beyond the Fock state and coherent state of light [50-53].

**Remarks on 'classical' spontaneous emission** - Here we address some queries regarding spontaneous emissions in classical and quantum theories. In a semiclassical model in which the expectation value of the squared absolute value of the wavefunction is interpreted as the density of the QEW [20], the solution of Maxwell equations results in wavepacket size and shape dependent spontaneous radiation, and in the case of density modulation – superradiant harmonic frequency emission, similar to the classically derived coherent radiation emission from a bunched point-particle beam of electrons in a variety of interaction schemes, such as Cherenkov, Smith-Purcell and Undulator-Synchrotron radiations [47,48]. However, the QED formulation of the same radiative interaction scheme [14] does not produce this wavepacket-dependent "spontaneous emission", and it produces only the manifestly wavepacket-independent expression (16) for both finite-size and modulated QEW. This quantum spontaneous emission originates from the "zero-point quantum vibration" and is the same as in the case of a plane-wave electron quantum wavefunction [5]. Thus, the QED analyses of the radiation emission from a single QEW fail to produce the "point particle" "classical spontaneous emission" limit [3,27, 34] (contrary to the stimulated interaction case).

It is our conjecture that the generalized formulation presented here, and particularly the new energy reciprocity relation (ARC) (4, 11), that originates from not neglecting the quantum interference terms (7, 8) would be instrumental in bridging the QED to classical theories of spontaneous emission from charged particles. These relations have wide validity, including for the case of entangled electron photon states (on which we did not expand in the present work).[63] We conject that further development of this methodologically new analysis of QEW radiation, would help to bridge the classical point-particle (probabilistic) current and the QED "zero-point vibration" presentations of spontaneous emission. Moreover, the ARC reciprocity relation seems to provide a new QEW approach for the investigation of the long-standing problem of single-particle Lorentz-Abraham self-force and radiation reaction effect in the second quantization treatment. [47, 64].

**Conclusion**

The main result of this work is the acceleration/radiation correspondence (ARC) (4, 11) that intimately connects the physics of particle acceleration and photon radiation at the fundamental level. These fundamental relations can only be derived in the framework of QED theory, and are derived through a methodologically new analysis of electron wave-interaction as a scattering process by keeping the interference terms between initial and scattered states, which have been previously ignored in scattering analyses based on Fermi's golden rule. Beyond showing consistency of our analysis with classical and semiclassical theories of phase and shape-dependent QED interaction with light at the coherent state, we presented new extension of the QED interaction theory to the case of interactions with quantum light – Fock state and squeezed state.

Note that our general new QED formalism beyond Fermi's golden rule allows for the case of entangled electron-light states that we did not explore in the present work. We conject that further development of this formulation can be useful in the study of interesting fundamental quantum theory problems, such as an electron wavefunction Schrodinger cat, the Lorentz-Abraham self-force and QED derivation of classical limit spontaneous emission, We raise the question whether in the wider sense, particle-wave duality beyond the Fermi Golden rule can also play a role in elementary particle interaction problems [65] and in macroscopic quantum/classical scattering processes such as time-periodic Schrodinger's equation in nonlinear optical processes [66].

### Acknowlegements

We acknowledge W. Schleich, P. Kling, Itay Griniasty, Yaron Silberberg for useful discussions and comments. The work was supported in parts by DIP (German-Israeli Project Cooperation) No. 04340302000, ISF (Israel Science Foundation) No. 00010001000 and by ICORE— Israel Center of Research Excellence program of the ISF, and by the Crown Photonics Center. Correspondance and requests for materials should be addressed to Y.P. (yiming.pan@weizmann.ac.il) or A. G.(gover@eng.tau.ac.il).

# Appendix A

To derive the photon emission and electron acceleration expression (19, 24), the integration over p in (9,10) should be carried out with the Gaussian distribution function of the drifted electron amplitude in momentum space. For the phase-independent second order photon emission (e) terms $\Delta v_q^{(2)}, \Delta E^{(2)}$, this involves the following integrations

$$\sum_p \left\{ \left( \frac{p + p_{rec}^{(e)} - \hbar q_z/2}{p_0} \right)^2 \left| c_{p+p_{rec}^{(e)}}^{(0)} \right|^2 \right\}$$

$$= \left( 2\pi\sigma_{p_0}^2 \right)^{-1/2} \int dp \left( \frac{p+p_{rec}^{(e)} - \hbar q_z/2}{p_0} \right)^2 \exp\left( -\frac{\left(p+p_{rec}^{(e)} - p_0\right)^2}{2\sigma_{p_0}^2} \right)$$

$$= \left( 1 - \frac{\hbar q_z}{2p_0} \right)^2 + \left( \frac{\sigma_{p_0}}{p_0} \right)^2 \approx 1,$$

$$\sum_p \left\{ \left( \frac{p + p_{rec}^{(e)} - \hbar q_z/2}{p_0} \right)^2 p \left| c_{p+p_{rec}^{(e)}}^{(0)} \right|^2 \right\}$$

$$= \left( 2\pi\sigma_{p_0}^2 \right)^{-1/2} \int dp \left( \frac{p+p_{rec}^{(e)} - \hbar q_z/2}{p_0} \right)^2 p \exp\left( -\frac{\left(p+p_{rec}^{(e)} - p_0\right)^2}{2\sigma_{p_0}^2} \right) \quad \text{(A1)}$$

$$\approx p_0.$$

Similarly, for the absorption (a) term

$$\sum_p \left\{ \left( \frac{p - (p_{rec}^{(a)} - \hbar q_z/2)}{p_0} \right)^2 \left| c_{p-p_{rec}^{(a)}}^{(0)} \right|^2 \right\} = \left( 1 + \frac{\hbar q_z}{2p_0} \right)^2 + \left( \frac{\sigma_{p_0}}{p_0} \right)^2 \approx 1,$$

$$\sum_p \left\{ \left( \frac{p - (p_{rec}^{(a)} - \hbar q_z/2)}{p_0} \right)^2 p \left| c_{p-p_{rec}^{(a)}}^{(0)} \right|^2 \right\} \approx p_0. \quad \text{(A2)}$$

For the phase-dependent first-order photon emission (e) part ($\Delta v_q^{(1)}, \Delta E^{(1)}$)

$$\sum_{p}\left\{\left(\frac{p+(p_{rec}^{(e)}-\hbar q_z/2)}{p_0}\right)\Re\left\{c_p^{(0)*}c_{p+p_{rec}^{(e)}}^{(0)}\right\}\right\}$$

$$=\left(2\pi\sigma_{p_0}^2\right)^{-1/2}\int dp\left(\frac{p+p_{rec}^{(e)}-\hbar q_z/2}{p_0}\right)\Re\left\{\exp\left(-\frac{(p-p_0)^2}{4\sigma_{p_0}^2(1-i\xi t_D)^{-1}}\right)\exp\left(-\frac{\left(p+p_{rec}^{(e)}-p_0\right)^2}{4\sigma_{p_0}^2(1+i\xi t_D)^{-1}}\right)\right\}$$

$$=e^{-\Gamma^2/2}, \tag{A3}$$

$$\sum_{p}\left\{\left(\frac{p+(p_{rec}^{(e)}-\hbar q_{zm}/2)}{p_0}\right)\Re\left\{c_p^{(0)*}c_{p+p_{rec}^{(e)}}^{(0)}\right\}\right\}(p-p_0)\approx-\left(\frac{\hbar\omega}{v_0}\right)e^{-\Gamma^2/2}.$$

Similarly, for the absorption (a) term

$$\sum_{p}\left\{\left(\frac{p-(p_{rec}^{(a)}-\hbar q_z/2)}{p_0}\right)\left(c_p^{(0)*}c_{p-p_{rec}^{(a)}}^{(0)}\right)\right\}=e^{-\Gamma^2/2}\left(1-\frac{p_{rec}^{(e)}-\hbar q_z+ip_{rec}^{(e)}\xi t_D}{2p_0}\right)\approx e^{-\Gamma^2/2},$$

$$\sum_{p}\left\{\left(\frac{p-(p_{rec}^{(a)}-\hbar q_z/2)}{p_0}\right)\left(c_p^{(0)*}c_{p-p_{rec}^{(a)}}^{(0)}\right)\right\}(p-p_0)\approx\left(\frac{\hbar\omega}{v_0}\right)e^{-\Gamma^2/2}, \tag{A4}$$

where we define the decay parameter $\Gamma=\left(\frac{\omega}{v_0}\right)\sigma_z(t_D)=\left(\frac{\hbar\omega}{v_0}\right)\frac{\sqrt{1+\xi^2 t_D^2}}{2\sigma_{p_0}}=\Gamma_0\sqrt{1+\xi^2 t_D^2}$ and $\Gamma_0=\frac{2\pi}{\beta}\left(\frac{\sigma_z}{\lambda}\right)\xi=2\sigma_{p_0}^2/m^*\hbar$ (19). Note that in all cases we took the approximation $p_{rec}^{(e,a)},\hbar q_z,\sigma_{p_0}\ll p_0$.

**Appendix B**

In an open structure, the general spectral radiant energy W per unit angle $d\Omega$ per frequency $d\omega$ is given by [6]

$$\left(\frac{d^2W}{d\Omega d\omega}\right)_{SP}\simeq\hbar\omega\rho_{ph}(\omega)\Delta v_{q,vac}, \tag{B1}$$

where spontaneous emission is $\Delta v_{q,vac}$ at the initial photon number $v_0=0$ and the free-space photon density of state is

$$\rho_{ph}(\omega)=\omega^2 V/8\pi^2 c^3. \tag{B2}$$

With $\Delta\nu_{q,\text{vac}} = \tilde{\Upsilon}^2 \text{sinc}^2(\bar{\theta}/2)$ (16), $\tilde{\Upsilon} = e\tilde{E}_{qz}L/4\hbar\omega$, and the quantization condition $\frac{1}{2}\sqrt{\frac{\varepsilon_0}{\mu_0}}|\tilde{E}_{q\perp 0}|^2 A_{\text{eff}} t_I = \hbar\omega$, it has been shown [14] that the Smith-Purcell spontaneous photon emission per steradian and unit frequency of a single electron is:

$$\left(\frac{d\nu}{d\omega d\Omega}\right)_{SP} = \frac{e^2 L^2}{64\pi^2} \frac{\omega^2}{c^2} \sqrt{\frac{\mu_0}{\varepsilon_0}} |\eta_{qm}|^2 \text{sinc}^2\left(\frac{\bar{\theta}}{2}\right) \quad (B3)$$

where $\eta_{qm} = E_{qzm}/E_{q\perp 0}$ is the fraction of the synchronous Floquet space harmonic amplitude relative to the fundamental, and $\bar{\theta}$ is the detuning parameter for $q_z = q_{zm} = (\omega/c)\cos\Theta + m2\pi/\lambda_G$.